\documentclass[twocolumn,showpacs,preprintnumbers,amsmath,amssymb]{revtex4}
\usepackage{graphicx}% Include figure files
\usepackage{dcolumn}% Align table columns on decimal point
\usepackage{bm}% bold math
\begin{document}
\title{Pauli spin susceptibility of a strongly correlated two-dimensional electron liquid}
\author{A.~A. Shashkin$^*$, S. Anissimova, M.~R. Sakr$^\dag$, and
S.~V. Kravchenko}
\affiliation{Physics Department, Northeastern University, Boston,
Massachusetts 02115, U.S.A.}
\author{V.~T. Dolgopolov}
\affiliation{Institute of Solid State Physics, Chernogolovka, Moscow
District 142432, Russia}
\author{T.~M. Klapwijk}
\affiliation{Kavli Institute of Nanoscience, Delft University of
Technology, 2628 CJ Delft, The Netherlands}
\begin{abstract}
Thermodynamic measurements reveal that the Pauli spin susceptibility
of strongly correlated two-dimensional electrons in silicon grows
critically at low electron densities --- behavior that is
characteristic of the existence of a phase transition.
\end{abstract}
\maketitle

Presently, theoretical description of interacting electron systems is
restricted to two limiting cases: (i)~weak electron-electron
interactions (small ratio of the Coulomb and Fermi energies
$r_s=E_C/E_F\ll1$, high electron densities) and (ii)~very strong
electron-electron interactions ($r_s\gg1$, very low electron
densities). In the first case, conventional Fermi-liquid behavior
\cite{landau} is established, while in the second case, formation of
the Wigner crystal is expected \cite{wigner} (for recent
developments, see Ref.~\cite{tanatar}). Numerous experiments
performed in both three- (3D) and two-dimensional (2D) electron
systems at intermediate interaction strengths ($1\lesssim
r_s\lesssim5$) have not demonstrated any significant change in
properties compared to the weakly-interacting regime (see, {\it
e.g}., Refs.~\cite{3d,2d}). It was not until recently that
qualitative deviations from the weakly-interacting Fermi liquid
behavior (in particular, the drastic increase of the effective
electron mass with decreasing electron density) have been found in
strongly correlated 2D electron systems ($r_s\gtrsim10$)
\cite{review1}. However, these findings have been based solely on the
studies of a kinetic parameter (conductivity), which, in general, is
not a characteristic of a state of matter.

The 2D electron system in silicon turns out to be a very convenient
object for studies of the strongly correlated regime due to the large
interaction strengths ($r_s>10$ can be easily reached) and high
homogeneity of the samples estimated (from the width of the
magnetocapacitance minima in perpendicular magnetic fields) at about
$4\times10^9$~cm$^{-2}$ \cite{shashkin01}. In this Letter, we report
measurements of the thermodynamic magnetization and density of states
in such a low-disordered, strongly correlated 2D electron system in
silicon. We concentrate on the metallic regime where conductivity
$\sigma\gg e^2/h$. We have found that in this system, the spin
susceptibility of band electrons (Pauli spin susceptibility) becomes
enhanced by almost an order of magnitude at low electron densities,
growing critically near a certain critical density
$n_\chi\approx8\times10^{10}$~cm$^{-2}$: behavior that is
characteristic in the close vicinity of a phase transition. The
density $n_\chi$ is coincident within the experimental uncertainty
with the critical density $n_c$ for the zero-field metal-insulator
transition (MIT) in our samples. The nature of the low-density phase
($n_s<n_\chi$) still remains unclear because even in the cleanest of
currently available samples, it is masked by the residual disorder in
the electron system.

Measurements were made in an Oxford dilution refrigerator on
low-disordered (100)-silicon samples with peak electron mobilities of
3~m$^2$/Vs at 0.1~K and oxide thickness 149~nm. These samples are
remarkable by the absence of a band tail of localized electrons down
to electron densities $n_s\approx1\times10^{11}$~cm$^{-2}$, as
inferred from the coincidence of the full spin polarization field
obtained from parallel-field magnetotransport and from the analysis
of Shubnikov-de~Haas oscillations (the former is influenced by
possible band tail of localized electrons, while the latter is not;
for more details, see Refs.~\cite{review1,moments,jp}). This allows
one to study properties of a {\em clean} 2D electron system without
admixture of local moments \cite{moments,jp,mott}. The second
advantage of these samples is a very low contact resistance (in
``conventional'' silicon samples, high contact resistance becomes the
main experimental obstacle in the low density/low temperature limit).
To minimize contact resistance, thin gaps in the gate metalization
have been introduced, which allows for maintaining high electron
density near the contacts regardless of its value in the main part of
the sample.

For measurements of the magnetization, the parallel magnetic field
$B$ was modulated with a small ac field $B_{\text{mod}}$ in the range
0.01 -- 0.03~T at a frequency $f=0.45$~Hz, and the current between
the gate and the two-dimensional electron system was measured with
high precision ($\sim10^{-16}$~A) using a current-voltage converter
and a lock-in amplifier. The imaginary (out-of-phase) current
component is equal to $i=(2\pi fCB_{\text{mod}}/e)\,d\mu/dB$, where
$C$ is the capacitance of the sample and $\mu$ is the chemical
potential. By applying the Maxwell relation $dM/dn_s=-d\mu/dB$, one
can obtain the magnetization $M$ from the measured $i$. A similar
technique has been applied by Prus {\it et al}.~\cite{prus03} to a 2D
electron system in silicon with high level of disorder, in which case
the physics of local moments has been mainly studied. As discussed
below, the data analysis and interpretation is not quite correct in
Ref.~\cite{prus03}; in particular, Prus {\it et al}.\ do not
distinguish between the Pauli spin susceptibility of band electrons
and the Curie spin susceptibility of local moments.

For measurements of the thermodynamic density of states, a similar
circuit was used with a distinction that the gate voltage was
modulated and thus the imaginary current component was proportional
to the capacitance. Thermodynamic density of states $dn_s/d\mu$ is
related to magnetocapacitance via $1/C=1/C_0+1/Ae^2(dn_s/d\mu)$,
where $C_0$ is the geometric capacitance and $A$ is the sample area.

\begin{figure}
\scalebox{0.36}{\includegraphics[clip]{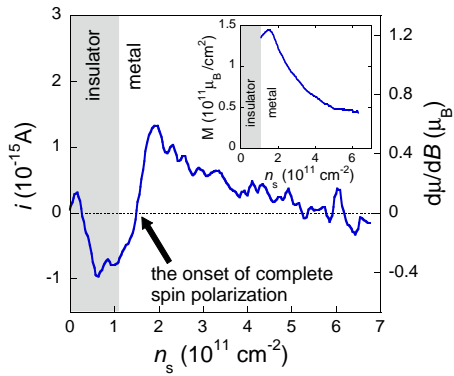}}
\caption{\label{fig1} Imaginary current component in the
magnetization experiment as a function of the electron density in a
magnetic field of 5~T and $T=0.4$~K. Grey area depicts the insulating
phase. Magnetization versus $n_s$ is displayed in the inset. Note
that the maximum $M$ is coincident within the experimental
uncertainty with $\mu_Bn_s$.}
\end{figure}

A typical experimental trace of $i(n_s)$ in a parallel magnetic field
of 5~T is displayed in Fig.~\ref{fig1}. The inset shows magnetization
$M(n_s)$ in the metallic phase obtained by integrating
$dM/dn_s=-d\mu/dB$ with the integration constant $M(\infty)=B\chi_0$,
where $\chi_0$ is the Pauli spin susceptibility of non-interacting
electrons. A nearly anti-symmetric jump of $i(n_s)$ about zero on the
$y$-axis (marked by the black arrow) separates the high- and
low-density regions in which the signal is positive and negative
($M(n_s)$ is decreasing and increasing), respectively. Such a
behavior is expected based on simple considerations. At low
densities, all electrons are spin-polarized in a magnetic field, so
for the simple case of non-interacting 2D electrons one expects
$d\mu/dB=-\mu_B$ (at $n_s\rightarrow0$, deep in the insulating
regime, the capacitance of the system vanishes and, therefore, the
measured current approaches zero). At higher densities, when the
electrons start to fill the upper spin subband, $M(n_s)$ starts to
decrease, and $d\mu/dB$ is determined by the renormalized Pauli spin
susceptibility $\chi$ and is expected to decrease with $n_s$ due to
reduction in the strength of electron-electron interactions. Finally,
in the high-density limit, the spin susceptibility approaches its
``non-interacting'' value $\chi_0$, and $d\mu/dB$ should approach
zero. The onset of complete spin polarization --- the electron
density $n_p$ at which the electrons start to fill the upper spin
subband --- is given by the condition $d\mu/dB=0$ ($M(n_s)$ reaches a
maximum), as indicated by the black arrow in the figure. It is
important that over the range of magnetic fields used in the
experiment (1.5--7~tesla), the maximum $M$ coincides within the
experimental uncertainty with $\mu_Bn_s$ thus confirming that all the
electrons are indeed spin-polarized below $n_p$. Note however that
the absolute value of $d\mu/dB$ at $n_s\lesssim n_c$ is reduced in
the experiment. We attribute this to smearing of the minimum in
$i(n_s)$ caused by possible influence of the residual disorder in the
electron system, which is crucial in and just above the insulating
phase, in contrast to the clean metallic regime we focus on here.
Another reason for the reduction in $d\mu/dB$ is the
electron-electron interactions (due to, {\it e.g.}, the enhanced
effective mass).

\begin{figure}
\scalebox{0.48}{\includegraphics[clip]{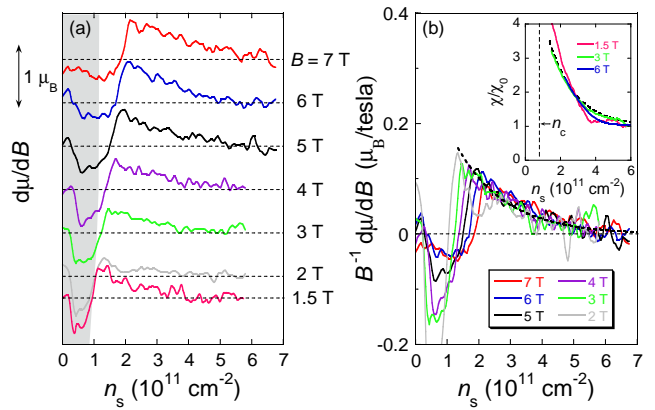}}
\caption{\label{fig2} (a)~The experimental $d\mu/dB$ as a function of
electron density in different magnetic fields and $T=0.4$~K. The
curves are vertically shifted for clarity. Grey area depicts the
insulating phase. Note that the onset of full spin polarization in
our experiment always takes place in the metallic regime. (b)~Scaling
of the $d\mu/dB$ curves, normalized by magnetic field magnitude, at
high electron densities. The dashed line represents the ``master
curve''. Spin susceptibility obtained by integrating the master curve
(dashed line) and raw data at $B=$~1.5, 3, and 6~T is displayed in
the inset.}
\end{figure}

In Fig.~\ref{fig2}(a), we show a set of curves for the experimental
$d\mu/dB$ versus electron density in different magnetic fields.
Experimental results in the range of magnetic fields studied do not
depend, within the experimental noise, on temperature below 0.6~K
(down to 0.15~K which was the lowest temperature achieved in this
experiment). The onset of full spin polarization shifts to higher
electron densities with increasing magnetic field. Grey area depicts
the insulating phase, which expands somewhat with $B$ (for more on
this, see Ref.~\cite{MIT}). Note that the range of magnetic fields
used in our experiment is restricted from below by the condition that
$d\mu/dB$ crosses zero in the metallic regime. In Fig.~\ref{fig2}(b),
we show how these curves, normalized by magnetic field, collapse in
the partially-polarized regime onto a single ``master curve''. The
existence of such scaling verifies proportionality of the
magnetization to $B$, confirming that we deal with Pauli spin
susceptibility of band electrons, and establishes a common zero level
for the experimental traces. Integration of the master curve over
$n_s$ yields the spin susceptibility $\chi=M/B$, as shown in the
inset to Fig.~\ref{fig2}(b). Also shown is the spin susceptibility
obtained by integration of raw curves at $B=$~1.5, 3, and 6 tesla,
which, within the experimental error, yield the same dependence.

This method of measuring the spin susceptibility, being the most
direct, suffers, however, from possible influence of the unknown
diamagnetic contribution to the measured $d\mu/dB$, which arises from
the finite width of the 2D electron layer \cite{rem}. To verify that
this influence is negligible in our samples, we employ another two
independent methods to determine $\chi$. The second method is based
on marking the electron density $n_p$ at which $d\mu/dB=0$ and which
corresponds to the onset of complete spin polarization, as mentioned
above. The so-determined polarization density $n_p(B)$ can be easily
converted into $\chi(n_s)$ via $\chi=\mu_Bn_p/B$. Note that in
contrast to the value of $d\mu/dB$, the polarization density $n_p$ is
practically not affected by possible influence of the diamagnetic
shift.

\begin{figure}
\scalebox{0.7}{\includegraphics[clip]{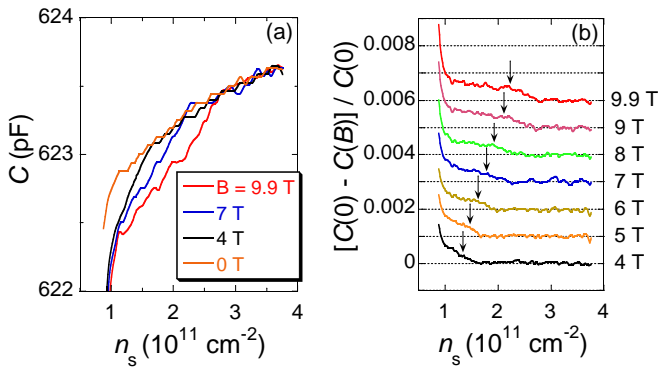}}
\caption{\label{fig3} (a)~Magnetocapacitance versus electron density
for different magnetic fields. (b)~Deviation of the $C(n_s)$
dependences for different magnetic fields from the $B=0$ reference
curve. The traces are vertically shifted for clarity. The onset of
full spin polarization is indicated by arrows.}
\end{figure}

The third method for measuring $n_p$ and $\chi$, insensitive to the
diamagnetic shift, relies on analyzing the magnetocapacitance, $C$.
Experimental traces $C(n_s)$ are shown in Fig.~\ref{fig3}(a) for
different magnetic fields. As the magnetic field is increased, a
step-like feature emerges on the $C(n_s)$ curves and shifts to higher
electron densities. This feature corresponds to the thermodynamic
density of states abruptly changing when the electrons' spins become
completely polarized. To see the step-like feature more clearly, in
Fig.~\ref{fig3}(b) we subtract the $C(n_s)$ curves for different
magnetic fields from the reference $B=0$ curve. The fact that the
jumps in $C$ (as well as in $d\mu/dB$) are washed out much stronger
than it can be expected from possible inhomogeneities in the electron
density distribution (about $4\times10^9$~cm$^{-2}$
\cite{shashkin01}) points to the importance of electron-electron
interactions. Since the effects of interactions are different in the
fully- and partially-polarized regimes, it is natural to mark the
onset of full spin polarization at the beginning of the
interaction-broadened jump, as indicated by arrows in the figure. In
case the residual disorder does contribute to the jump broadening, we
extend error bars to the middle of the jump, which yields an upper
boundary for the onset of full spin polarization.

\begin{figure}
\scalebox{0.33}{\includegraphics[clip]{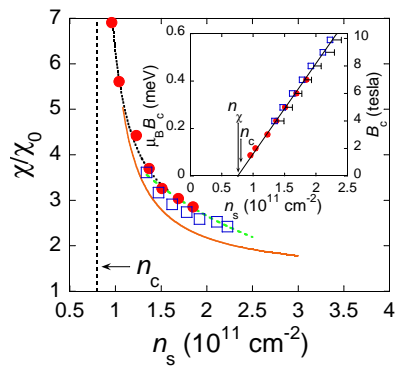}}
\caption{\label{fig4} Dependence of the Pauli spin susceptibility on
electron density obtained by all three methods described in text:
integral of the master curve (dashed line), $d\mu/dB=0$ (circles),
and magnetocapacitance (squares). The dotted line is a guide to the
eye. Also shown by a solid line is the transport data of
Ref.~\cite{shashkin01}. Inset: polarization field as a function of
the electron density determined from the magnetization (circles) and
magnetocapacitance (squares) data. The symbol size for the
magnetization data reflects the experimental uncertainty, and the
error bars for the magnetocapacitance data extend to the middle of
the jump in $C$. The data for $B_c$ are described by a linear fit
which extrapolates to a density $n_\chi$ close to the critical
density $n_c$ for the $B=0$ MIT.}
\end{figure}

In Fig.~\ref{fig4}, we show the summary of the results for the Pauli
spin susceptibility as a function of $n_s$, obtained using all three
methods described above. The excellent agreement between the results
obtained by all of the methods establishes that a possible influence
of the diamagnetic shift is negligible \cite{remark} and, therefore,
the validity of the data including those at the lowest electron
densities is justified. There is also good agreement between these
results and the data obtained by the transport experiments of
Ref.~\cite{shashkin01}. This adds credibility to the transport data
and confirms that full spin polarization occurs at the field $B_c$;
however, we note again that evidence for the phase transition can
only be obtained from thermodynamic measurements. The magnetization
data extend to lower densities than the transport data, and larger
values of $\chi$ are reached, exceeding the ``non-interacting'' value
$\chi_0$ by almost an order of magnitude. The Pauli spin
susceptibility behaves critically close to the critical density $n_c$
for the $B=0$ metal-insulator transition \cite{rem1}: $\chi\propto
n_s/(n_s-n_\chi)$. This is in favor of the occurrence of a
spontaneous spin polarization (either Wigner crystal \cite{rem2} or
ferromagnetic liquid) at low $n_s$, although in currently available
samples, the formation of the band tail of localized electrons at
$n_s\lesssim n_c$ conceals the origin of the low-density phase. In
other words, so far, one can only reach an incipient transition to a
new phase.

The dependence $B_c(n_s)$, determined from the magnetization and
magnetocapacitance data, is represented in the inset to
Fig.~\ref{fig4}. The two data sets coincide and are described well by
a common linear fit which extrapolates to a density $n_\chi$ close to
$n_c$. We emphasize that in the low-field limit ($B<1.5$~tesla), the
jump in $d\mu/dB$ shifts to the insulating regime, which does not
allow us to approach closer vicinity of $n_\chi$: based on the data
obtained in the regime of strong localization, one would not be able
to make conclusions concerning properties of a clean metallic
electron system which we are interested in here. Clearly, the fact
that the linear $B_c(n_s)$ dependence persists down to the lowest
electron densities achieved in the experiment confirms that we always
deal with the clean metallic regime.

Finally, we would like to clarify the principal difference between
our results and those of Ref.~\cite{prus03}. In the sample used by
Prus {\it et al}., the critical density $n_c$ for the $B=0$ MIT was
considerably higher than in our samples caused by high level of
disorder, and the band tail of localized electrons was present at all
electron densities \cite{prus03}. As a result, the crucial region of
low electron densities, in which the critical behavior of the Pauli
spin susceptibility occurs, falls within the insulating regime where
the physics of local moments dominates \cite{moments,jp,mott}.
Indeed, Prus {\it et al}.\ have found sub-linear $M(B)$ dependence
characteristic of local moments, and the extracted spin
susceptibility in their sample has a Curie temperature dependence
\cite{jp}. This is the case even at high electron densities, where
metallic behavior might be expected instead. Such effects are absent
in our samples: the spin susceptibility (in the partially-polarized
system) is independent of the magnetic field and temperature,
confirming that we deal with Pauli spin susceptibility of band
electrons.

In summary, the Pauli spin susceptibility has been determined by
measurements of the thermodynamic magnetization and density of states
in a low-disordered, strongly correlated 2D electron system in
silicon. It is found to behave critically near the zero-field MIT,
which is characteristic of the existence of a phase transition.

We gratefully acknowledge discussions with S. Chakravarty, D. Heiman,
N.~E. Israeloff, R.~S. Markiewicz, and M.~P. Sarachik. One of us
(SVK) would like to thank B.~I. Halperin for suggesting this method
to measure spin susceptibility. We would also like to thank A.
Gaidarzhy and J.~B. Miller for technical assistance and C.~M. Marcus
and P. Mohanty for an opportunity to use their microfabrication
facilities. This work was supported by NSF grant DMR-0403026, PRF grant 41867-AC10, the RFBR, RAS, and the Programme ``The State Support of Leading Scientific Schools''.


\begin{thebibliography}{apssamp}
\bibitem{rem} To deal with this problem, the ``subtraction of the
diamagnetic contribution'' was suggested in Ref.~\cite{prus03}. The
diamagnetic contribution $\Delta$ was determined in the
partially-polarized regime as the difference between the direct
magnetization data (``Mag'') and the data obtained from
Shubnikov-de~Haas oscillations (``SdH''):
$\Delta=\mbox{Mag}-\mbox{SdH}$. The experimental data were then
corrected by $\Delta$. We find this procedure meaningless as it
essentially results in replacing the magnetization data by
Shubnikov-de~Haas data:
$\mbox{Mag}-\Delta=\mbox{Mag}-(\mbox{Mag}-\mbox{SdH})=\mbox{SdH}$. In
fact, the difference between magnetization and Shubnikov-de~Haas data
in their experiment is likely to be due to the presence of a band
tail of localized electrons at all electron densities in their
sample.
\bibitem{remark} Since the diamagnetic shift decreases with
increasing $n_s$ and/or decreasing $B$, it may in principle be
noticeable at the lowest $n_s$ and highest $B$ used in the
experiment. Comparing low- and high-field curves shown in
Fig.~\ref{fig2}(a), we estimate that at
$n_s\sim1\times10^{11}$~cm$^{-2}$ and $B=7$~T, the contribution of
the diamagnetic shift is less than $0.2\mu_B$.
\bibitem{rem1} The critical density for the MIT was determined from
transport measurements (see Refs.~\cite{review1,MIT}).
\bibitem{rem2} The fact that $B_c\rightarrow0$ as $n_s\rightarrow
n_c$ speaks in favor of the strongly correlated liquid being close to
the crystal \cite{nozieres79}.
\end{thebibliography}
\end{document}